*Epistemological remarks about the Ghirardi-Rimini-Weber quantum theory.*

**Francesco de Stefano, retired teacher of Mathematics and Physics**

Liceo Scientifico "G. Marinelli" – Udine – Italy

fdestefano55@gmail.com

## Abstract

We discuss the epistemological features of the Ghirardi-Rimini-Weber (GRW) proposal to modify quantum mechanics (QM) in order to overcome the objectification problem, that means the transition between the quantum and the classical level of description of natural phenomena. This proposal also solves the so called "quantum measurement problem" (QMP) together with some of the most intruiging quantum paradoxes, such as Schrödinger's cat and Wigner's friend ones. We shall show that the GRW is a "realistic" theory, but in a way quite different from "Einstein's realism" or other kinds of realism proposed in the past, such as the Many World Interpretation (MWI) or the Bohmian Mechanics (BM).

*Introduction*

In 1986[1] G.C. Ghirardi, A. Rimini and T. Weber advanced a proposal to overcome the conceptual and epistemological problems of the Quantum Theory of Measurement (QTM) and in particular the well known problem of the transition between the quantum and the classical description of the physical world. This last topic is also known as the "objectification problem"[2], that means that, at the classical level, a quantum superposition of macroscopically distinguished states is impossible. As it is well known, at the microscopic level the effect of the entanglement is unavoidable, meaning that a composite system of two (or more) parts which have interacted in the past loses all the properties of the individual parts[3] (quantum non-separability). Only the whole system has properties compatible with the Heisenberg uncertainty principle. And, moreover, when a measurement on one of the parts is performed, individual properties appear also on the part non subjected to the process of measurement: this is the non-locality effect of QM so well discussed and underlined by J. S. Bell[4]. These non-locality effects are now very well established by hundreds of experiments[5] and nearly nobody denies them. Of course, these intriguing features of QM gave origin to a widespread epistemological debate about the "reality" in the quantum world: the well known and celebrated Einstein-Bohr debate is only the emerging part of this iceberg[6]! Nevertheless, both Einstein and Bohr agreed about the fact that, at the macroscopic level, the entanglement does not appear, and every macroscopic object is never in a superposition of distinguished states (that means, that in every moment it has specific and "real" properties, observer-independent). But how and where is the transition between the quantum and the classical level? Where is the border between them? When does it happen? And, finally but very important, which is the precise mechanism of this transition? All these topics were widely discussed during the whole past century, and the subtle analysis of Bell played a central role in this debate[7]. These discussions led also to deep philosophical and epistemological debates about the concept of "reality" in the physical world[8].

The aim of this paper is to analyze the proposal of GRW as a solution of the objectification problem, as an answer to the questions about the quantum/classical transition and about a realistic approach to QM: a realism of course compatible with the quantum nonlocality. I will try to show that the GRW theory is, in my opinion, the best approach, up to now, to solve the above quoted problems and also the best solution to the "arrow of time" one: the emergence of the irreversibility of the classical world from the reversibility of the physical laws at the microscopic level.

*A short review of the GRW*

In the beginning, the aim of GRW was simply to build up a model in order to overcome the conceptual problems of the QTM and of the transition between the quantum and the classical levels of the description of the physical world. The purpose of the three italian physicists was essentially to establish a well defined border between these two levels. In fact, from the most celebrated description of this problem by J. von Neumann[9], the impossibility to precisely define this border has been very well known[10]. And strictly connected to this topic was also the objectification problem, which was very well underlined by the very famous Schrödinger's cat paradox[11]. Let us summarize, in a very short way, the objectification problem inside the QTM. Consider a system **S** which is in a superposition of quantum states (to simplify, only two, like the singlet state of the spin) $|\psi\rangle = c_1|\alpha_1\rangle + c_2|\alpha_2\rangle$ where $|\alpha_1\rangle$ and $|\alpha_2\rangle$ are the eigenstates of a quantum observable $\mathcal{A}$ and $|c_1|^2$ and $|c_2|^2$ the corresponding probabilities (according to the Born's rule) to find the eigenvalues $a_1$ or $a_2$ when a measurement of $\mathcal{A}$ is performed. Send now **S** into a measuring apparatus **A** which can measure this observable; of course, at the beginning it is in a state $|A_0\rangle$ which tells that it has not just measured the value of $\mathcal{A}$. Then, the composite system of **S** + **A** will be described by the quantum state $|\psi\rangle|A_0\rangle$[1]. Then, according to the unitary quantum evolution (Schrödinger equation), the system will evolve into the entangled state $|\psi'\rangle = c_1|\alpha_1\rangle|A_1\rangle + c_2|\alpha_2\rangle|A_2\rangle$ where **A** is now in a superposition of macroscopically distinguishable states. In some sense, we can say that its index "dances" between the two values $a_1$ and $a_2$. Or, better, that the index of the apparatus is in a "sum state" which does not mean that it is in one of them, or in both of them or in no one of them (the only logical alternatives). This is the core of the objectification problem: according to the standard QM, not only **S** has not a definite value for the observable $\mathcal{A}$, but even the measuring apparatus is not in a definite state. But, has anybody seen such an apparatus? Of course not! Then the Copenhagen interpretation of QM simply avoids the problem introducing the famous wave packet reduction postulate (WPR), telling that, after the measurement is performed, the system **S** + **A** collapses, totally random and with the probability previously quoted, into one of the terms of the entangled state, $|\alpha_1\rangle|A_1\rangle$ or $|\alpha_2\rangle|A_2\rangle$.

This was the state of the art from the birth of the quantum formalism described by von Neumann till 1964 (with the great intermediate step of the Einstein-Podolsky-Rosen argument[6]. In that year the very famous paper by Bell[4] appeared and this opened a great and long phase of deep developments of QM in order to solve this problem, avoiding the "ad hoc" (in Popper's language) argument of the WPR postulate. The GRW model of 1986 is one of these developments of the standard QM. In this paper I will refer to the original work of GRW: in fact, in order to analyze the epistemological features of this proposal, it is not necessary neither consider the CSL model[12] developed together with P. Pearle and R. Grassi, nor the so called GRWm theory[13] which has completed and deeply developed the initial one. Then, what is the central idea of the GRW theory? It is based on one fundamental concept: while the wavefunction of a system evolves according the ordinary Schrödinger equation, it is "sometimes" subjected to a "spontaneous localization", which gives it the "objective property" of the position. And this process is totally stocastical and non-linear: that is exactly what is necessary to preserve, on one hand, the non epistemic role of the quantum probabilities and, on the other hand, the impossibility of the existence of superposition of macroscopically distinguished states for a measuring apparatus or, more in general, for a classical system (like... a cat!)

---

[1] In the whole paper we will refer to the case of an "ideal measurement", that is a measurement which does not change the state of S. This is not a restriction, in principle, to the conceptual problems of QTM. It would be easy to show that all that we will tell is tenable (see ref. 3).

which has interacted with the quantum one. Bell was very strongly impressed by the GRW proposal and not only wrote to Ghirardi an important letter to express his appreciation[14], but even wrote, in his paper for Schrödinger's centennial[15], a specific chapter with the title "Ghirardi, Rimini and Weber". Following Bell's analysis, we can say that the main idea of GRW is that while a wavefunction $\psi(t;\vec{r}_1,\vec{r}_2,....,\vec{r}_N)$ evolves according to the usual Schrödinger equation, it is subjected to a "jump", a spontaneous localization, with the probability $\frac{N}{\tau}$, where N is the number of the arguments $\vec{r}$ in the wavefunction and $\tau$ is a time constant, which Bell himself considered a "new constant of nature", and not simply a "parameter" like GRW called it. This jump leads to a new wavefunction $\psi' = \frac{j(\vec{x}-\vec{r}_k)\psi}{R_k(\vec{x})}$ where $\vec{r}_k$ is randomly chosen among the arguments $\vec{r}$ in the wavefunction, and $j(\vec{x}-\vec{r}_k)$ is the "jump function", normalized. For this jump function, GRW suggest a Gaussian one $j(\vec{x}) = K\exp\left(-\frac{\vec{x}^2}{2a^2}\right)$ where $a$ is another parameter (considered by Bell as a second new constant of Nature). The theory then strongly depends on these two constants and GRW propose a reasonable choice of their values: $\tau \approx 10^{15}\, s \approx 10^8\, years$ and $a \approx 10^{-5}\, cm$. This means that for a single particle a spontaneous localization occured very seldom up to now in the history of the Universe, that means that the "normal" evolution of the quantum particle is widely coincident with the unitary evolution of the Schrödinger equation. But, even if very very little, it exists a difference from the GRW probability predictions and those coming from the Born rule. And this, in Bell's opinion, was a relevant feauture of the GRW theory, because it is experimentally testable[16]. But, while in a quantum system the GRW evolution and the Schrödinger's one are pratically equivalent, when we consider a great system (like a measuring apparatus or… a cat!), the spontaneous localization is amplified by the number of particles (typically $10^{23}$, the Avogadro number), that means that the mean lifetime before a GRW jump becomes $\frac{10^{15}}{10^{23}} = 10^{-8}\, s$ !

This implies that the measuring apparatus or the Schrödinger's cat do not remain in a superposition of macroscopically distinguishable states for more than such a time. Then the pointer of the apparatus shows a very definite outcome of the experiment and the cat is definitively life or dead. This very precise mathematical border was, in Bell's opinion, the most important and interesting feature of the GRW proposal. Quoting directly Bell's words

> "For myself, I see the GRW model as a very nice illustration of how quantum mechanics, to become rational, requires only a change which is very small (on some measures!). And I am particularly struck by the fact that the model is as Lorentz invariant as it could be in the nonrelativistic version. It takes away the ground of my fear that any exact formulation of quantum mechanics must conflict with fundamental Lorentz invariance"

After this short review of the original GRW model, I will make the epistemological analysis of it and its deep connection to the philosophical problem of compatibility of QM with a realist conception of the world.

***GRW and philosophical realism***

Let us come to the philosophical and epistemological features of the GRW theory. But first of all, I want to underline some developments of the original GRW model which have led to the actual and more sophisticated theory. Of course, as well as Bell remarked[15], one of the open questions in the model was its relativistic extension. This was the main development of the theory and was in the beginning faced by GRW with the very important contribution of Grassi and Pearle. This cooperation had as a result the so called CSL model (Continous Spontaneous Localization), which is a more precise and deep version of the GRW paper of 1986[12]. It is out of the topic of this paper to analyze the CSL, mainly because it does not imply deeper remarks at the epistemological level, whose analysis is my main purpose.

A further development of the theory is the so called GRWm model (where m stands for "mass") in which an important role is represented by a "mass density" involved in the mechanism of the spontaneous localization. Finally I want to quote an important contribution of R. Tumulka[17] who has made important steps towards the relativistic version of the GRW theory.

And now, let us enter into the philosophical debate.

First of all, we can ask ourselves why GRW choose the "position" as the most relevant observable in the theory. In fact, the spontanous localization mechanism concerns about the position of a particle. We can make two remarks: 1) from the physical point of view, it is very important to underline that every experiment or every statement about a physical property deals with "positions" of something: the position of the index of an analogical apparatus, giving the outcome of the experiment, or the position of a flash of a photon or a particle on a screen, or the position of the leds on a digital display, giving the value found; 2) from the philosophical point of view, every sentence about the "reality" of something also deals with its position: this table is here or there, John Fitzgerald Kennedy was killed in Dallas on November 22nd 1963, the pen on the table is blue… Of course, I have written "reality" because we will see that just the definition of realism is the most intruiging and questionable topic, from the epistemological point of view. But, in order to begin, this rough and intuitive interpretation of realism will be sufficient[2].

It is also very important to tell that in the GRW theory all the most charateristic feuatures of QM are preserved: the entanglement, what Schrödinger himself called "the fundamental trait of QM", is valid (at least for $10^{15}$ years, as we have seen above); the non-locality (the most important behaviour of the quantum world, as pointed out by Bell) occurs; the non epistemic probabilities (that means that the quantum probabilities are not a matter of our ignorance about the details of a quantum system, but they are intrinsically ontological) are present.

In order to discuss now the epistemology of this theory, it is unavoidable to enter in a deeper way in the different definitions of realism that we can consider relevant for this debate in QM. To this purpose, I will refer to a very deep and interesting paper by T. Norsen[8] that I discovered many years ago and suggested to Ghirardi, who was deeply impressed by it and stimulated very important discussions and considerations. In this paper, Norsen wants to underline the numerous misunderstandings about Bell's theorem: many authors in fact considered that in Bell's analysis a very important role is played by the concept of realism and introduced this problem mixing it to the non-locality of QM. But Norsen showed that this is a wrong interpretation of Bell's work: tho only and important feature underlined by Bell is the quantum non-locality, totally independent from any conception of realism that one can have in mind. In his paper, Norsen talks about four kinds of realism: *naive, scientific, perceptual and metaphysical realism*.

---

[2] It is important to remark that the most relevant Hidden Variables Theory (HVT), the BM, considers the position as the main physical variable which is in every moment well specified for a particle. All the other physical observable are contextual[18].

Quoting Norsen... "In the philosophy of perception, Naive Realism is the view that all features of a perceptual experience have their origin in some identical corresponding feature of the perceived object. For example, a Naive Realist will say that, when someone sees a red apple, the experienced redness resides in the apple, as a kind of intrinsic property that is passively revealed in the perceptual experience."

It is obvious, considering what we have told about the GRW, that this theory does not adapt to this definition: in fact when a spontaneous localization occurs, according to QM, the considered particle in general has not the "intrinsic property" of the position and this appears with the localization, in a totally random and epistemical probability. Naive realism is typical of positivism and of a great part of classical physics (from Galileo to the beginning of XIX century).

Let us consider now scientific realism. In Norsen's words "In the philosophy of science, this is the doctrine that we can and should accept well-established scientific theories as providing a literally-true description of the world." As it would be clear, this kind of realism is a step forward the naive one. In fact, naive realism taks about only the perceptual experiences, while scientific realism deals with the scientific theories as a whole. Of course, scientific realism contains naive realism (the objectivity of the perceptions) but it concerns also the explanations of the connections between the perceptions (the theories). In fact, scientific realism is commonly opposed to the "instrumentalism", an epistemological point of view for which the theories are simply connections between the perceptions and are only instruments to this goal. Inside instrumentalism we can not speak about a true or a false theory, but only make declarations about their utility or not. Personally, I do not believe that the GRW theory is a kind of this realism: Ghirardi (who was my supervisor in the Master of Physics and my good friend for nearly 40 years[14]) had a very deep epistemological culture. His philosophical view was very near to Thomas Kuhn's one[19]. In Kuhn's epistemology we can speak about the existence of a progress in science, but inside the concept of "paradigm", which has not an ontological validity in itself and then the scientific theories inside the paradigm are not true (or false) in the meaning of the scientific realism, but only more or less coherent with the paradigm. This shifts the problem of truth to the paradigms. But in Kuhn's view a paradigm can not be judged as true or false, because the paradigms are incommensurable: this means that a paradigm conqueres the dominant position inside the scientific commmunity mainly through "extra-scientific" reasons. Then we can talk about a progress in science only because a new paradigm (after a scientific revolution, in Kuhn's words) has brought to a deeper and better description of the physical world. But what is "deeper and better" is still up to now a matter of discussion around Kuhn's philosophy of science. In every case, I can say that the GRW approach is not realist in the sense of the scientific realism described by Norsen.

The third type of realism listed by Norsen, is perceptual realism. In his words "...I will mean the idea that sense perception provides a primary and direct access to facts about the world – i.e., that what we are aware of in normal perception is *the world*, and not any sort of subjective fantasy, inner theater, or mental construction." An immediate question arises: which is the difference between perceptual and naive realism? I have read many times Norsen's paper and, in my opinion, it is really very difficult to find a significant difference between the two kinds of realism. Norsen remarks that perceptual realism presupposes the existence of a "real world" outside us (a conception that we will see more embedded in metaphysical realism) and our perceptions lead directly to it. Maybe in the naive version of realism we only talk about the single perceptions and do not make a particular affirmation about the existence of a global world which is at the bottom of our perceptions. Perceptual realism then makes a step more. In every case, I think that both kinds of realism conflict, in a strong way, with what the modern neurosciences tell us about how the brain builds what we call "knowledge of the world"[20,21,22,23]. It is commonly recognized now that what we call *knowledge* is not a simple "input-output" mechanism, in which we receive informations and inputs from Nature and then we produce, as an output, our theories which describe Nature as well as possible. Even the process of visual perception is more dependent from what happens inside our brain than from what comes into our brain! In fact only the 20% of a visual perception is given by the external contribution of it: the rest comes from our previous knowledge, our memory, and the internal connections that the structures of the brain make[21]. It seems that Norsen does not take into account these features of the so called "biology of the knowledge", which is, in my opinion, the new frontier of epistemology. In every case I think that also perceptual realism is not at the basis of the GRW epistemology. Ghirardi (at least from our private discussions) had not a specific competence in these neuroscientific topics but was deeply aware of them and surely had a "constructivist" position from the epistemological point of view.
Then let us look at the last kind of realism quoted by Norsen: the metaphysical realism.
Following Norsen "This Realism accepts the existence of an external world, but without necessarily requiring anything specific in regard to its similarity to the world of our perceptual experience or the account of any particular scientific theory." This kind of realism is also the philosophical choice of Norsen himself. And this is really a metaphysical position, taking into account that this "act of faith" is independent from any particular scientific theory describing this world and even any specific kind of sensitive perception in which we can believe. Of course, this definition of realism seems to be acceptable by everyone who wants to work in a scientific field. In other words, it would be very strange for a physicist or a biologist to produce a research program if they would not believe in the existence of an external world, whose existence is independent from the presence of human beings on the Earth. Obviously, the epistemological problem of "how" we have knowledge of this world, "which" are the scope and limits of this knowledge, "if" this knowledge is progressive or not, is a matter of choice of the specific epistemogical attitude that one assumes. For example, Norsen remarks that even the Many Worlds Interpretation (MWI) of QM is realist in this last sense. Then I can say that this is the most general defintion of realism we can deal with: even the celebrated "reality criterion" present in the Einstein-Podolsky-Rosen[6] (EPR) paper is more strict.
Then, surely the GRW theory is a realist theory in this perspective. But also the MWI and the BM are so. Then, is this simply the end of the game? No, of course. After having established that the GRW is a realistic theory (in the sense of metaphysical realism) I will underline now why, in my opinion, it is the most promising approach to solve the conceptual puzzles of QM. First of all, as we have seen, it solves the objectification problem of the QTM, forbidding the superposition of macroscopically distinguished states for more than $10^{-5}$ s. And this well defined border between the microscopic and the macroscopic levels is experimentally testable (even if up to now not yet made): then its realism, at this level, is stronger than MWI, which is not experimentally testable. Secondly, it solves all the paradoxes like the Schrödinger's cat or the Wigner's friend, from the moment that the cat is in a superposition of states "dead" and "alive" for not more than the same little time interval quoted before. It would be interesting to see if the GRW permits, for example, "the Coronavirus of GRW", that means a superposition of different state for this virus: this would be also an interesting test of macrorealism like that implied by the Leggett-Garg inequality[24]. Finally, the GRW theory can be a very interesting solution to the problem of the "arrow of time", that is the emergence of the irreversibility of the macroscopic world from the reversibility of the microscopic one. I can justify this with two considerations. The first is that the GRW theory modifies the Schrödinger equation (which is linear and leads to a reversible behaviour) with a stocastic and non-linear term. This means that, in some sense, the seeds of the irreversibility are present from the beginning. But a second remark is more relevant. As it was shown by D. Albert[25], if we consider a thermodynamical system composed by two bodies with a difference of temperature among them, and we put them into contact, of course the system will reach an equilibrium state at an intermediate temperature. Statistical Mechanics (SM), that is the usual tool to describe thermodynamics, tells us that this happens because the final state has a higher probability occurence than the initial one. This means that there are more microstates compatible with the macrostate of equilibrium than with the two bodies with different temperatures. At this point Albert remarks that the usual explanation coming from SM is the same coming from the standard QM: but if we look at the example with a better attention, we see that, in the ensamble of the microstates belonging to a macrostate, there are "good" and "bad" microstates, that means microstates which are compatible or not with the evolution towards the equilibrium macrostate. But the number of the good microstates is larger than the number of the bad ones, and it is sufficient a little perturbation of the macrostate in order to bring it towards the set of the good microstates. And what is the origin of this kind of perturbation? Albert answers: the spontaneous localization in the GRW model! Then, if the GRW theory would be positively experimentally tested, the theory would automathically also explain the emergence of the irreversibility and then of the second principle of thermodynamics and then of the arrow of time!

*Conclusion*

I hope that these epistemological remarks can be shared by the community which adheres to the GRW proposal to modify QM. Every comment and correction is welcome. Quoting Karl Raimund Popper… "Research never ends"!